\documentstyle[aps,preprint,epsfig]{revtex}

\begin{document}

\author{N.\ Bulut, T.\ Dahm and D.J.\ Scalapino\\ }

\title{Monte Carlo study of the pairing interaction \\
in the two-leg Hubbard ladder}

\address{ 
Department of Physics, University of California \\
Santa Barbara, CA 93106-9530}

\maketitle
\begin{abstract}
Monte Carlo calculations of the irreducible particle-particle
interaction on a two-leg Hubbard ladder doped near 
half-filling are reported.
As the temperature is lowered, this interaction develops 
structure in momentum space similar to the magnetic 
susceptibility $\chi({\bf q})$ and reflects the 
development of strong short-range antiferromagnetic correlations.
Using this interaction, the eigenfunction of the leading singlet
pair eigenvalue is found to have $d_{x^2-y^2}$ like symmetry.
The single-particle spectral weight is also shown to peak near
$(\pi,0)$ and $(0,\pi)$ when the ratio of the inter- to 
intra-chain hopping $t_{\perp}/t\simeq 1.5$, 
leading to an increased tendency for pairing.
\end{abstract}


Numerical calculations, renormalization-group bosonization studies
as well as strong-coupling treatments find that half-filled $t$-$J$
or Hubbard two-leg ladders have a spin gapped ground state with 
short range antiferromagnetic 
correlations \cite{Rice,Noack,Balents}.
Furthermore, these various techniques all find that when holes
are doped into the ladder,
$d_{x^2-y^2}$-like pairing correlations develop.
This being the case, one would like to understand the nature 
of the effective interaction that gives rise to the $d_{x^2-y^2}$
pairing correlations for this system.
Here, in order to address this question, we 
present Monte Carlo results for the irreducible particle-particle 
interaction $\Gamma_{\rm I}$.
In addition, we use $\Gamma_{\rm I}$ and the single-particle Green's 
function obtained from the Monte Carlo calculations
to solve the Bethe-Salpeter equation in the singlet pairing channel.
We find that the eigenfunction with the leading 
eigenvalue has $d_{x^2-y^2}$-like symmetry.

The Hubbard model Hamiltonian for a two-leg ladder has the form
\begin{equation}
H = -t \sum_{i,\lambda,s}
\left( c^\dagger_{i\lambda s} c_{i+1\lambda s} + {\rm h.c.}
\right) - 
t_\perp \sum_{i,s} \left( c^\dagger_{i1 s}c_{i2 s} + {\rm
h.c.}\right) +
U \sum_{i\lambda} n_{i\lambda\uparrow} n_{i\lambda\downarrow}.
\label{eq:H}
\end{equation}
Here $t$ is the intra-chain one electron hopping, 
$t_{\perp}$ the inter-chain hopping and $U$ 
the on-site Coulomb interaction.
The operators $c_{i\lambda s}^{\dagger}$ and 
$c_{i\lambda s}$ create and destroy electrons of spin $s$ 
on site $i$ of the $\lambda$$^{\rm th}$ leg, 
respectively, and 
$n_{i\lambda s}=c_{i\lambda s}^{\dagger}c_{i\lambda s}$ is the 
occupation number for spin $s$ on site $i$ of the 
$\lambda$$^{\rm th}$ leg.

Using Monte Carlo techniques, we have calculated the finite 
temperature two-particle Green's function
\begin{equation}
G_2(x_4,x_3,x_2,x_1) =
-\langle \, T\,
c_{\uparrow}(x_4) c_{\downarrow}(x_3)
c^{\dagger}_{\downarrow}(x_2) c^{\dagger}_{\uparrow}(x_1) \rangle.
\label{G2}
\end{equation}
Here $c_s^{\dagger}(x_i)$ with 
$x_i=({\bf x}_i,\tau_i)$ 
creates an electron of spin $s$ at site ${\bf x}_i$ 
and imaginary time $\tau_i$ and $T$ is the usual 
$\tau$-ordering operator.
Then, as previously discussed \cite{vertex}, 
one can take the Fourier transform 
of both the space and imaginary time variables and obtain
$G_2(p',k',k,p)$ with 
$p'=({\bf p}',i\omega_{n'})$, etc.
This two particle Green's function can be expressed in terms of the
exact single-particle propagator $G_s({\bf p},i\omega_n)$ 
and the reducible particle-particle vertex $\Gamma(p',k',k,p)$ 
\begin{eqnarray}
G_2(p',k',k,p)\, & & = 
-\delta_{p,p'}\,\delta_{k,k'} \,
G_{\downarrow}(k) G_{\uparrow}(p) \\
\label{G22}
& & 
+ {T\over N}
\delta_{p'+k',p+k}\,
G_{\uparrow}(p') G_{\downarrow}(k')
\Gamma(p',k',k,p)
G_{\downarrow}(k) G_{\uparrow}(p). \nonumber
\end{eqnarray}
Then from the Monte Carlo data for $G$ and $G_2$, 
one can determine $\Gamma(p',k',k,p)$.
Finally, because the effective pairing interaction corresponds
to the irreducible particle-particle interaction $\Gamma_{\rm I}$
in the zero energy and momentum center of mass channel,
we have inverted the fully dressed $t$-matrix
equation to find $\Gamma_{\rm I}$
in terms of $\Gamma$ and $G$.
Setting $k=-p$ and
$k'=-p'$,
the fully dressed $t$-matrix equation becomes
\begin{equation}
\Gamma(p'|p)= \Gamma_{\rm I}(p'|p) 
-{T \over N} \sum_{k}
\Gamma(p'|k) 
G_{\downarrow}(-k) G_{\uparrow}(k)
\Gamma_{\rm I}(k|p),
\label{t-matrix}
\end{equation}
which is solved to find $\Gamma_{\rm I}(p'|p)$.
The effective pairing interaction in the singlet channel is 
\begin{equation}
V(p'-p) = {1 \over 2}
\Big(\Gamma_{\rm I}(p'|p)+\Gamma_{\rm I}(-p'|p)\Big).
\label{V}
\end{equation}
We have carried out this calculation at different temperatures
for various values of the hopping anisotropy $t_{\perp}/t$,
interaction strength $U/t$ and filling 
$\langle n\rangle=\langle n_{i\uparrow}+n_{i\downarrow}\rangle$.
Here, we will show results for $t_{\perp}/t = 1.5$,
$U/t=4$, $\langle n\rangle =0.875$
and a $2\times 16$ lattice.
We have chosen this set of parameters,
because the numerical density matrix renormalization group (DMRG)
calculations find that for $U/t=4$ and $\langle n\rangle = 0.875$,
the $d_{x^2-y^2}$ pairing correlations are strongest when 
$t_{\perp}/t\simeq 1.5$ \cite{DMRG}.
In addition, 
for these parameters we have good control of the 
maximum entropy analytic continuation of the Monte Carlo data,
which is necessary to obtain the single-particle spectral weight 
$A({\bf p},\omega)$.

\begin{figure}
\centerline{\epsfysize=7.5cm \epsffile[18 184 592 598]
{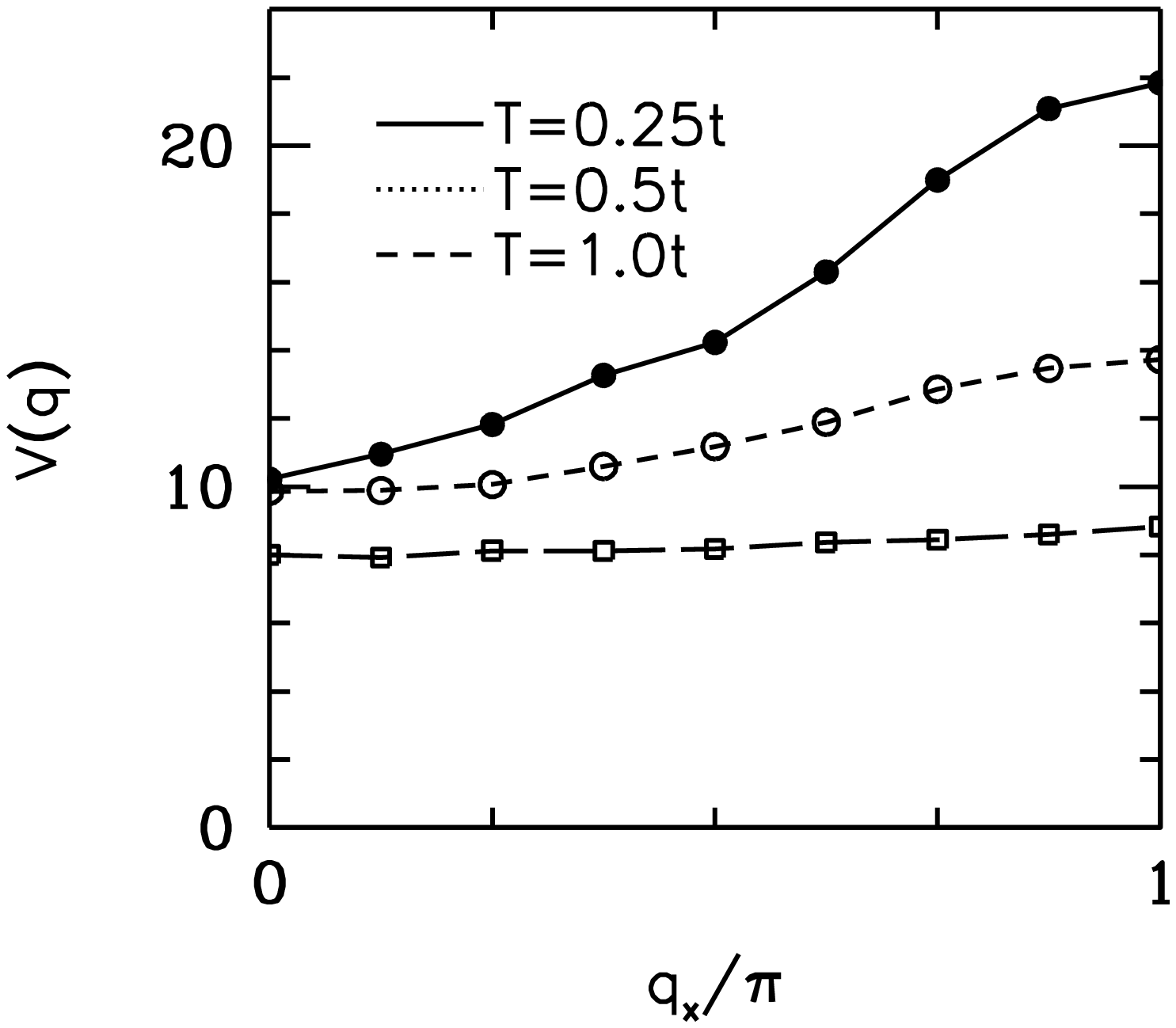}}
\vskip 1cm
\caption{
Momentum dependence of the effective interaction 
$V({\bf q})$ for 
$U=4t$, $\langle n\rangle=0.875$ and $t_{\perp}=1.5t$.
Here $V({\bf q})$ is measured in units of $t$,
$q_y=\pi$ and $V({\bf q})$ is plotted as a function of 
$q_x$.
}
\label{fig:Veff}
\end{figure}

\begin{figure}
\centerline{\epsfysize=7.5cm \epsffile[18 184 592 598]
{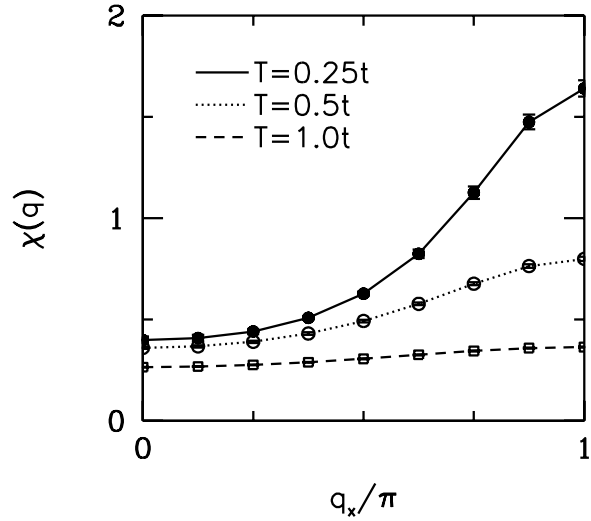}}
\vskip 1cm
\caption{
Momentum dependence of 
the magnetic susceptibility $\chi({\bf q})$ for 
$U=4t$, $\langle n\rangle=0.875$ and $t_{\perp}=1.5t$.
Here
$q_y=\pi$ and $\chi({\bf q})$ is plotted as a function of 
$q_x$.
}
\label{fig:chiq}
\end{figure}

In Fig. \ref{fig:Veff}
we plot the effective pairing interaction $V$ versus
$q_x=p_x'-p_x$
for $q_y=p_y'-p_y=\pi$.
Here we have set $\omega_n=\omega_{n'}=\pi T$ 
corresponding to $\omega_m=0$ energy transfer.
The three curves correspond to temperatures $T=1.0t$,
$0.5t$ and $0.25t$.
One can see that as the temperature decreases, the effective 
pairing interaction becomes increasingly positive at large 
momentum transfer ${\bf q}\rightarrow (\pi,\pi)$.
As shown in Figure \ref{fig:chiq}, the magnetic 
susceptibility 
\begin{equation}
\chi({\bf q}) = {1 \over N}
\int_0^{\beta}\,d\tau\,
\sum_{\ell} \,e^{i{\bf q}\cdot {\ell} } \,
\langle m_{i+\mathbf{\ell}}^-(\tau) m_i^+(0) \rangle
\label{eq:chiq}
\end{equation}
also develops structure at large momentum transfers over 
this same temperature region.
Thus it is clear that the effective pairing interaction is 
associated with the development of short-range antiferromagnetic
correlations.

\begin{figure}
\centerline{\epsfysize=7.5cm \epsffile[18 184 592 598]
{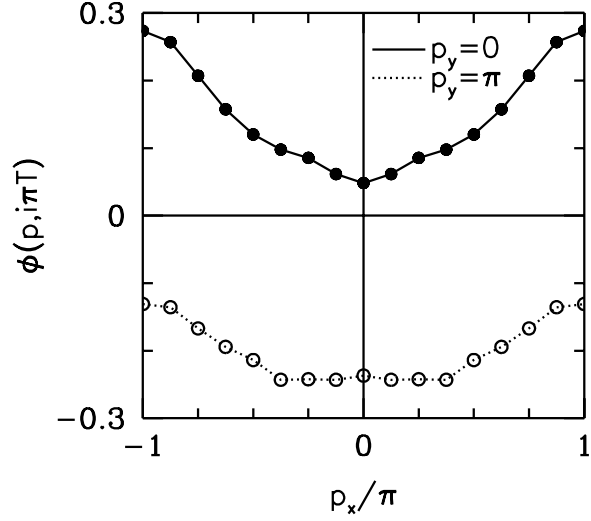}}
\vskip 1cm
\caption{
Momentum dependence of the $d_{x^2-y^2}$ eigenfunction
$\phi({\bf p},i\pi T)$.
These results are for $T=0.25t$,
$U=4t$, $\langle n\rangle=0.875$ and $t_{\perp}=1.5t$.
}
\label{fig:phi}
\end{figure}

It is also of interest to study the Bethe-Salpeter equation for the 
singlet particle-particle channel
\begin{equation}
\lambda_{\alpha} \phi_{\alpha}({\bf p},i\omega_n) = 
- {T\over N} \sum_{{\bf p}',i\omega_{n'}}
\Gamma_{\rm I}({\bf p},i\omega_{n}|{\bf p}',i\omega_{n'})
|G({\bf p}',i\omega_{n'})|^2
\phi_{\alpha}({\bf p}',i\omega_{n'}).
\label{BS}
\end{equation}
Fig. \ref{fig:phi} shows the eigenfunction $\phi({\bf p},i\omega_n)$
of the leading eigenvalue versus ${\bf p}$ for $\omega_n=\pi T$.
We observe that $\phi({\bf p},i\pi T)$ has 
$d_{x^2-y^2}$-like momentum structure in the sense that 
it has opposite signs and is largest near $(\pi,0)$
on the bonding band and near $(0,\pi)$ on the antibonding band.
This is associated with the structure of the irreducible 
interaction $\Gamma_{\rm I}$ and the single-particle spectral weight 
$A({\bf p},\omega)$ which we will study below \cite{note1}.

\begin{figure}
\centerline{
\epsfysize=7cm \epsffile[-10 184 562 598]{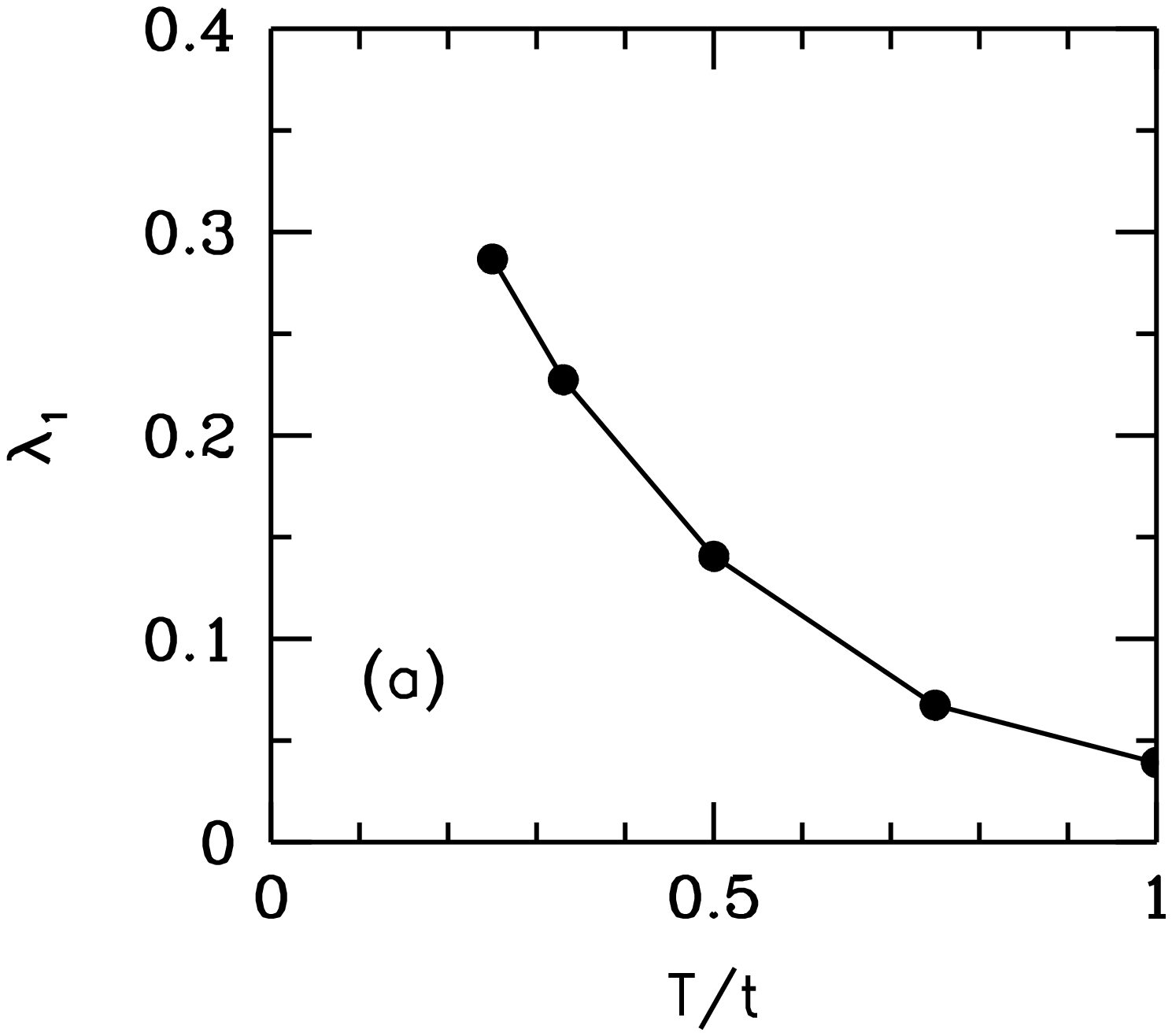}
\epsfysize=7cm \epsffile[78 184 652 598]{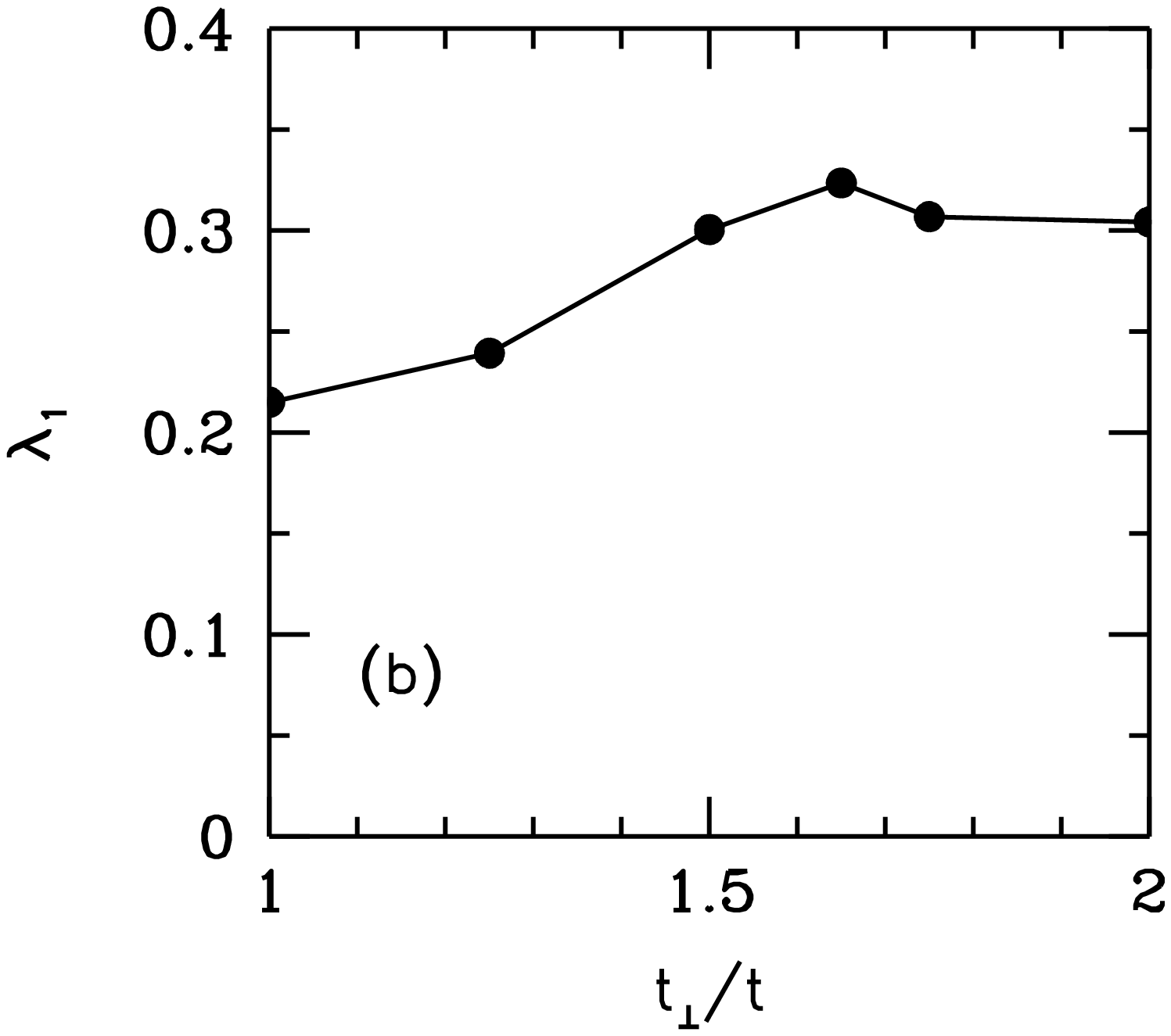}}
\vskip 1cm
\caption{
(a) Temperature dependence of the leading eigenvalue $\lambda_1$
for $t_{\perp}=1.5t$.
(b) $\lambda_1$ vs $t_{\perp}/t$ for 
$T=0.25t$.
These results are for $\langle n\rangle=0.875$ and $U=4t$.
}
\label{fig:lambda}
\end{figure}

The temperature dependence of the leading eigenvalue $\lambda_1$ 
is plotted in Fig. \ref{fig:lambda}(a).
As previously discussed, when the temperature is lowered,
short-range antiferromagnetic correlations develop and the effective 
pairing interaction increases at large momentum transfer.
This leads to an increase in $\lambda_1$ 
as shown in Fig. \ref{fig:lambda}(a).
We believe that $\lambda_1$ will approach unity at low temperatures
where power-law $d_{x^2-y^2}$-like pairing correlations have been
shown to exist using density matrix renormalization group 
techniques \cite{Noack}.
In Fig. \ref{fig:lambda}(b), we show the dependence of the leading eigenvalue
$\lambda_1$ on the hopping anisotropy $t_{\perp}/t$.
According to the DMRG calculations \cite{DMRG}, for $U/t=4$ 
and $\langle n\rangle=0.875$ the pairing correlations 
in the ground state are strongest when $t_{\perp}/t\simeq 1.5$.
In Fig. \ref{fig:lambda}(b), we do not observe a strong dependence of 
$\lambda_1$ on $t_{\perp}/t$ because of the thermal smearing 
effects at $T=0.25t$.

\begin{figure}
\vskip 1.5cm
\centerline{\epsfysize=7.5cm \epsffile[18 284 592 698]
{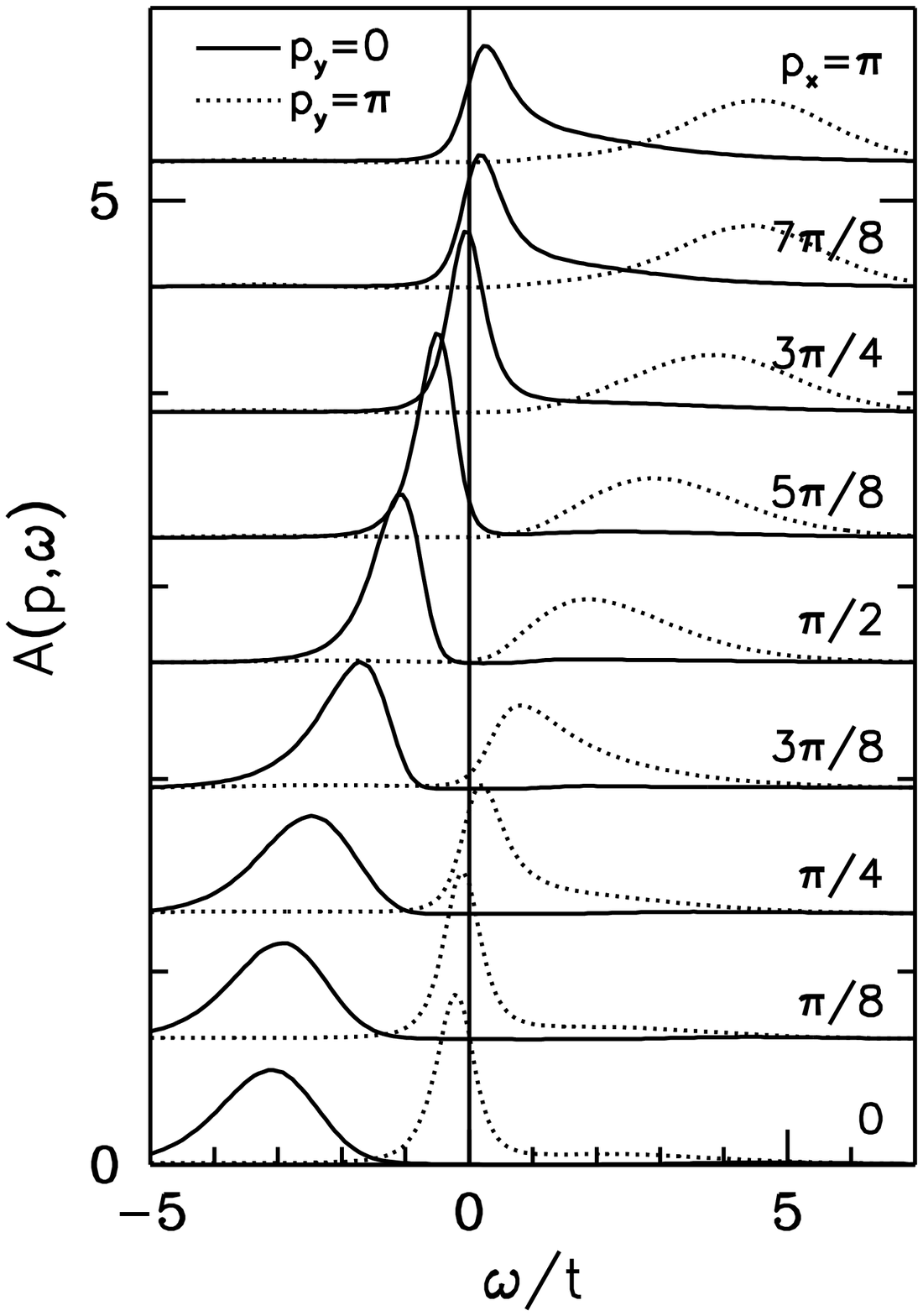}}
\vskip 3.5cm
\caption{
Single-particle spectral weight $A({\bf p},\omega)$ versus
$\omega$ for $t_{\perp}/t=1.5$, $T=0.25t$, $U/t=4$
and $\langle n\rangle=0.875$.
The solid curves denote the results for the bonding 
band ($p_y=0$) and the dotted curves denote the results 
for the antibonding band ($p_y=\pi$).
}
\label{fig:apw}
\end{figure}

In addition to the irreducible interaction vertex $\Gamma_{\rm I}$,
the single-particle Green's function $G({\bf p},i\omega_n)$ 
is also important in determining the structure of the 
leading eigenfunction of the Bethe-Salpeter equation. 
Using a numerical maximum entropy procedure \cite{White},
we have calculated the single-particle spectral weight 
\begin{equation}
A({\bf p},\omega)=
-{1\over \pi} {\rm Im}\,G({\bf p},\omega).
\label{Apw}
\end{equation}
This is plotted in Fig. \ref{fig:apw} 
for $t_{\perp}/t=1.5$ as a function of 
$\omega$ for different ${\bf p}$.
Here, the solid curves are for the bonding band ($p_y=0$) and the 
dotted curves are for the antibonding band ($p_y=\pi$).
We see that for this value of $t_{\perp}/t$, 
the bonding band has spectral weight near 
the Fermi level for ${\bf p}\sim (\pi,0)$, and the antibonding 
band has spectral weight near the Fermi level for 
${\bf p}\sim(0,\pi)$.
Hence, these Fermi points can be connected by scatterings
involving ${\bf q}=(\pi,\pi)$ momentum transfer.
Since $\Gamma_{\rm IS}$ is large and repulsive for 
${\bf q}\sim(\pi,\pi)$, the leading eigenfunction 
$\phi$ of the Bethe-Salpeter equation has
opposite signs for ${\bf p}$ near  
${\bf p}=(\pi,0)$ and $(0,\pi)$,
as seen in Fig. \ref{fig:phi}.

These calculations provide further insight into the structure 
of the effective pairing interaction and the single-particle 
spectral weight which lead to the pairing correlations 
in the two-leg Hubbard ladder.
Specifically, the momentum structure of the effective 
interaction $V({\bf q})$ clearly reflects the existence of 
short-range antiferromagnetic correlations as the cause of the 
increasing positive strength of $V({\bf q})$ at large momentum 
transfer.
Secondly, the enhanced spectral weight in the bonding band 
near $(\pi,0)$ and the antibonding band near $(0,\pi)$ are 
reminiscent of a similar effect observed in the two-dimensional
Hubbard model near half-filling \cite{BSW,Preuss}
and in the  ARPES of the cuprates \cite{Shen}.
This enhanced low lying spectral weight associated with the
renormalized quasiparticles is such that pair scattering processes 
with momentum transfers near $(\pi,\pi)$ have increased phase space.
These two features appear to play an important role in the
development of pairing on the two-leg ladder,
just as they do for the two-dimensional Hubbard model.

\acknowledgments

The authors gratefully acknowledge support from 
the National Science Foundation under Grant No. 
DMR95-27304, DMR95-20636 and PHY94-07194,
and from the 
Deutsche Forschungsgemeinschaft.
The numerical computations reported in this paper were performed 
at the San Diego Supercomputer Center.

\vfill
 
\vfill\eject

\begin{thebibliography}{999}
\bibitem{note1} We note that as $t_{\perp}/t$ is 
varied, the momentum structure of the single-particle spectral
weight and the irreducible interaction changes.
Consequently, the momentum structure of the leading eigenfunction
$\phi$ changes.
A discussion of how $\phi$ is affected by $t_{\perp}/t$ 
and further results on the doping dependence 
will be given in a longer article.
Here, we concentrate on a set of parameters for which the DMRG 
calculations find enhanced pairing correlations.
\end{thebibliography}
\end{document}